\documentclass[letter]{aa}

\usepackage{natbib}
\bibpunct{(}{)}{;}{a}{}{,}

\usepackage{graphicx}

\usepackage{txfonts}

\usepackage[]{hyperref}

\begin{document}

\title{One's loss is (not) another's gain}

\subtitle{Isotropic re-emission destabilizes mass transfer from radiative donor stars}

\author{
K. D. Temmink
\inst{1}
\fnmsep\thanks{e-mail: \href{mailto:Karel.Temmink@ru.nl}{Karel.Temmink@ru.nl} }
\and
S. Justham
\inst{2}
\and
O. R. Pols
\inst{1}
}

\institute{
Department of Astrophysics/IMAPP, Radboud University Nijmegen, P.O. Box 9010, 6500 GL Nijmegen, The Netherlands
\and
Max-Planck-Institut f\"{u}r Astrophysik, Karl-Schwarzschild-Stra{\ss}e 1, 85741 Garching, Germany
}
\date{Received XXXXXX; accepted YYYYYY}

\abstract
{Non-conservative mass transfer plays a central role in close-binary evolution, yet its effects on mass-transfer stability are uncertain. One widely adopted prescription, isotropic re-emission, is often assumed to promote stability compared to conservative mass transfer.}
{We investigate the impact of isotropic re-emission on the stability of mass transfer in binaries with radiative envelopes that undergo delayed dynamical instability (DDI). We assess whether simplified criteria used in binary population synthesis codes accurately capture stability boundaries under varying mass-transfer efficiencies.}
{We perform detailed stellar evolution calculations for a set of representative binaries undergoing DDI. Varying the mass-transfer efficiency $\beta$, we track the onset of instability and quantify the corresponding critical mass ratio. We compare our results with predictions from the commonly used $\zeta$-method, which is based on mass-radius exponents.}
{We find that a lower mass-transfer efficiency destabilizes mass transfer in DDI systems, whereas the $\zeta$-method predicts that isotropic re-emission should stabilize it. The discrepancy arises because the $\zeta$-method fails to capture the full evolution of the orbit and mass ratio during pre-instability mass transfer. In some cases, the critical mass ratio is underestimated by nearly a factor of two.}
{Our findings show that isotropic re-emission can reduce, rather than enhance, DDI stability, underscoring the limitations of using fixed critical mass ratios and $\zeta$-based criteria. This highlights the need for calibrated prescriptions that capture the time-dependent evolution of mass ratio and orbital separation, with direct implications for modelling X-ray binaries, symbiotic stars, and double white dwarfs, including their transient rates and delay-time distributions.}


\maketitle

\section{Introduction}

The stability of mass transfer is a key factor in determining the outcomes of close binary evolution. Stable mass transfer via Roche-lobe overflow can lead to long-lived semi-detached configurations, such as Algol-type binaries \citep{vanRensbergen2011}, as well as detached binaries in wider orbits after mass transfer ends (e.g., hot subdwarf + main-sequence systems; \citealt{Han2002}). Unstable mass transfer is thought to lead to a common-envelope (CE) phase \citep{Paczynski1976a, Ivanova2013b}, which can result in either a merger or the formation of a short-period binary \citep[e.g., compact hot subdwarf binaries;][]{Han2002,Heber2016}. Identifying the conditions that separate these outcomes is essential for understanding the formation of systems like X-ray binaries, double white dwarfs, or compact object mergers \citep{Webbink1984, Iben1985, Postnov2014}.

A central aspect of mass transfer stability is the response of the binary orbit to the redistribution and loss of mass. This response is influenced by both the efficiency of mass transfer (the fraction of the mass lost by the donor that is gained by the companion star) and the amount of angular momentum carried away by any material lost from the system \citep{Kalogera1996, Soberman1997, Tauris2006a}.

One commonly adopted assumption, usually referred to as isotropic re-emission, is that any mass lost from the system is expelled from the vicinity of the accreting star and carries away the accretor's specific orbital angular momentum. This prescription is physically motivated when the accretor cannot retain the transferred material, for example during super-Eddington transfer onto neutron stars or black holes (as in the context of SS433-like systems; \citealt{vandenHeuvel2017}). It is also widely adopted as a parameterised description of non-retained material in white-dwarf accretion contexts (e.g. in population-synthesis studies; \citealt{Claeys2014}). For these reasons, isotropic re-emission is frequently used in binary evolution and population synthesis models, including work on merging binary black-hole progenitors \citep[e.g.][]{Picco2024}.

When considering the instantaneous response of a binary to mass transfer, isotropic re-emission may be expected to favour stable mass transfer because the donor’s Roche lobe tends to contract less steeply per unit donor mass lost than in the fully conservative case \citep[e.g.][]{Soberman1997}. This can make it easier for the donor to remain within its Roche lobe. However, because the ejected material carries away orbital angular momentum, the orbit can still shrink strongly, and understanding how these competing effects play out in time-dependent evolutions remains an active topic.

In this study, we clarify the impact of isotropic re-emission on stability in systems that undergo delayed dynamical instability (DDI). In such cases, which arise for donors with radiative envelopes \citep{Hjellming1987}, mass transfer starts out stable but becomes unstable after substantial mass loss. We show that, when the time-dependent pre-instability evolution is taken into account, isotropic re-emission can induce instability after less donor mass has been lost than in the fully conservative case, and can furthermore destabilize systems that would remain stable under conservative mass transfer. We quantify this effect using detailed stellar evolution models.

This work is organized as follows: in Section~\ref{sec:AM_and_DDI} we review the analytical stability criteria used in binary population synthesis models, and describe how these relate to orbital evolution and angular momentum loss. We describe our numerical setup and binary models in Section~\ref{sec:param_space}, where we also present our results. Section~\ref{sec:implications_and_conclusions} discusses conclusions from our results.

\section{Isotropic re-emission in delayed dynamical instability binaries}
\label{sec:AM_and_DDI}

\begin{figure}[tb!]
\centering
\includegraphics[width=\hsize]{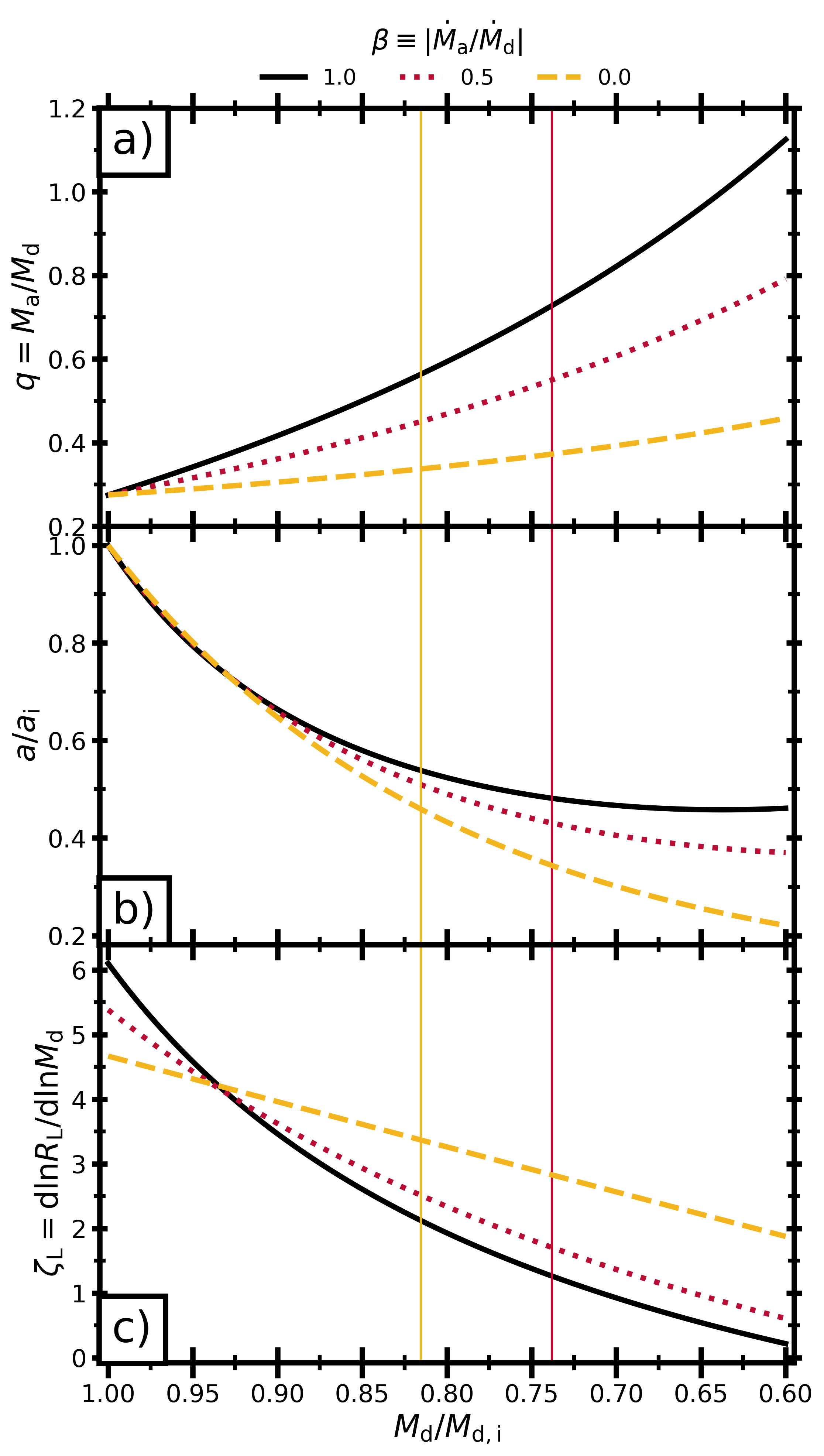}
\caption{Evolution of the mass ratio, semi-major axis and \(\zeta_{\rm L}\) computed using an analytical orbital-evolution prescription (Eqs.~\ref{eq:orbital_evolution} and~\ref{eq:zeta_lobe}), for an initial mass ratio \( M_{\rm a}/M_{\rm a} = 0.275\). The colours and styles of the curves correspond to different mass-transfer efficiencies, as indicated in the legend. Vertical lines indicate when the quasi-adiabatic instability criterion \citep[see][]{Temmink2023a} is reached in our MESA simulations of a binary with the same values of $\beta$ and initial mass ratio, an initial donor mass of 5\,$M_\odot$ and an initial period of about 7 days (see Appendix~\ref{sec:example_MESA_models}). Mass transfer is stable in the fully conservative case.}
\label{fig:illustrative_example}
\end{figure}

To characterize the evolution of mass-transferring binaries, we first briefly review the analytic orbital-evolution formalism describing how the orbit and donor star respond to variations in mass-transfer efficiency and angular momentum loss. We denote the binary mass ratio as \(q \equiv M_{\rm a} / M_{\rm d}\), where \(M_{\rm a}\) and \(M_{\rm d}\) are the accretor and donor masses, respectively. The mass-transfer efficiency is characterized by the parameter \(\beta \equiv \left | \dot{M}_{\rm a} / \dot{M}_{\rm d} \right |\), such that \(\beta = 1\) corresponds to fully conservative transfer and \(\beta = 0\) to complete re-ejection of the transferred material. In the isotropic re-emission mode considered here, the ejected mass is assumed to carry the specific orbital angular momentum of the accretor.

In binary evolution models, the stability of mass transfer is typically assessed using one of two approaches. The first method (hereafter called the $\zeta$-method) compares the donor's radial response to mass loss with that of its Roche lobe, assuming that the donor star responds adiabatically. The donor's response is described by $\zeta_{\rm d} = \mathrm{d} \ln R_{\rm d} / \mathrm{d} \ln M_{\rm d}$, and the Roche lobe's by $\zeta_{\rm L} = \mathrm{d} \ln R_{\rm L} / \mathrm{d} \ln M_{\rm d}$. If \(\zeta_{\rm d} > \zeta_{\rm L}\), the donor can remain within its Roche lobe upon rapid mass loss, and mass transfer is stable. If \(\zeta_{\rm d} < \zeta_{\rm L}\), however, the donor expands relative to its Roche lobe and the mass transfer is unstable. The second method (hereafter called the $q_{\rm crit}$-method) compares the mass ratio \(q = M_{\rm a} / M_{\rm d}\) to a critical threshold \(q_{\rm crit}\); if \(q < q_{\rm crit}\), mass transfer is expected to be dynamically unstable, whereas if \(q > q_{\rm crit}\), it proceeds stably.

These two criteria are equivalent in cases where mass transfer becomes unstable immediately upon Roche-lobe overflow (as is approximated for donors with deep convective envelopes), since \(q_{\rm crit}\) can be derived from the condition \(\zeta_{\rm d} = \zeta_{\rm L}\) if the mass-transfer efficiency and angular momentum loss prescription are specified \citep{Soberman1997, Tauris2006a}. However, the equivalence breaks down in systems that undergo DDI, where mass transfer begins stably but becomes dynamically unstable only after the donor has lost sufficient mass. In such cases, the \(\zeta\) method cannot account for the later onset of instability. In practice, population synthesis implementations that implement the \(\zeta\) method therefore adopt an effective donor response calibrated to reproduce the stability boundary seen in detailed calculations of systems undergoing DDI, rather than the donor’s instantaneous \(\zeta_{\rm d}\) \citep{Toonen2012, Riley2022}. It may nevertheless be regarded as more self-consistent than assuming a fixed \(q_{\rm crit}\). This is because changes in the mass-transfer efficiency and degree of angular momentum loss are reflected directly in the Roche-lobe response \(\zeta_{\rm L}\), whereas these dependencies are not naturally captured by fixed critical mass-ratio prescriptions. The \(q_{\rm crit}\) formalism can instead approximate the boundary between stable and unstable systems by incorporating the cumulative effects of mass and angular momentum loss. Neither method, however, can model the DDI process itself, including the duration of stable transfer and the amount of mass lost before instability sets in.

Assuming mass loss occurs exclusively through the isotropic re-emission mode with a constant $\beta$, the evolution of the binary semi-major axis $a$ may be calculated analytically as \citep{Kalogera1996, Soberman1997, Tauris1999}:
\begin{align}
\label{eq:orbital_evolution}
\frac{a}{a_{\rm i}} = \begin{cases} \displaystyle
\left( \frac{M_{\rm d}}{M_{\rm d, i}}\right)^{-2} \left( \frac{M_{\rm d} + M_{\rm a}}{M_{\rm d, i} + M_{\rm a, i}}\right)^{-1} \cdot e^{2(M_{\rm d} - M_{\rm d, i})/M_{\rm a}} & \text{if $\beta = 0$}\\
\displaystyle \left( \frac{M_{\rm d}}{M_{\rm d, i}}\right)^{-2} \left( \frac{M_{\rm a}}{M_{\rm a, i}}\right)^{-2/\beta} \left( \frac{M_{\rm d} + M_{\rm a}}{M_{\rm d, i} + M_{\rm a, i}}\right)^{-1} & \text{otherwise}.
\end{cases}
\end{align}
Here, the subscript $i$ denotes initial values. From this, in combination with the expression for $R_{\rm L}$ due to \citet{Eggleton1983}, it follows that
\begin{align}
\label{eq:zeta_lobe}
\zeta_{\rm L} &= [1 + \beta q^{-1}] \psi + (2 + 3\beta)q^{-1},
\end{align}
with
\begin{align*}
\psi &= -\frac{4}{3} - \frac{1}{1 + q} - \frac{2/5 + \frac{q^{1/3}}{3} (1 + q^{-1/3})^{-1}}{0.6 + q^{2/3}\ln (1 + q^{-1/3})}.
\end{align*}

These quantities, as well as the binary mass ratio, are shown in Fig.~\ref{fig:illustrative_example} for a representative binary with an initial mass ratio \( M_{\rm a}/M_{\rm a} = 0.275\). At the onset of Roche-lobe overflow, conservative mass transfer ($\beta = 1$) causes the orbit to contract most rapidly per unit donor mass lost (see panel b of Fig.~\ref{fig:illustrative_example}), despite the absence of angular momentum loss. This contraction is driven by changes in the mass ratio $q = M_{\rm a} / M_{\rm d}$: in the conservative case, the orbital angular momentum $J = \mu \sqrt{G (M_{\rm d} + M_{\rm a}) a}$ is constant, where $\mu = M_{\rm d} M_{\rm a}/(M_{\rm d} + M_{\rm a})$ is the reduced mass. Since $q$ and consequently $\mu$ increases (as initially $q <1$), the orbital separation $a$ must decrease to keep $J$ constant, and the orbit shrinks until the mass ratio reverses (i.e. $q=1$), after which it expands.

In non-conservative scenarios ($\beta < 1$), where some transferred mass escapes the system, the mass ratio $q = M_{\rm a} / M_{\rm d}$ evolves more gradually (see panel a of Fig.~\ref{fig:illustrative_example}) and the reduced mass $\mu = M_{\rm d} M_{\rm a} / (M_{\rm d} + M_{\rm a})$ changes more slowly with donor mass. At the same time, angular momentum is lost from the system, but the total binary mass $M_{\rm tot} = M_{\rm d} + M_{\rm a}$ also decreases, which partially offsets the effect of angular momentum loss by allowing a wider orbit for a given angular momentum. The orbital response is governed by a competition between these effects: angular momentum loss promotes shrinkage, while slower evolution with donor mass of $\mu$ and a decreasing $M_{\rm tot}$ act to preserve a wider orbit for a given angular momentum. At early stages of mass loss, the slower growth of $q$ typically dominates, and the orbit contracts less steeply with decreasing donor mass than in the conservative case. After substantial mass loss (at lower $M_{\rm d}$), however, the more extreme value of $q$ (compared to the fully conservative case) means that the orbit starts contracting more (per unit donor mass lost) than it would in a fully conservative case. Eventually, the orbit stops contracting and begins to expand. Although this transition occurs after a smaller change in $q$ than in the conservative case (see Appendix~\ref{sec:turning_point}), the donor must lose more mass to reach the minimum separation due to lower mass transfer efficiency.

Over the course of mass transfer, these differences accumulate. The slower evolution of $q$ also affects the Roche lobe's response, quantified by $\zeta_{\rm L} \equiv \mathrm{d}\ln R_{\rm L}/\mathrm{d}\ln M_{\rm d}$ (Eq.~\ref{eq:zeta_lobe}, panel c of Fig.~\ref{fig:illustrative_example}). In non-conservative models, where $q$ remains more extreme, $\zeta_{\rm L}$ changes more gradually. As a result, a point is eventually reached (after the donor has lost 7.5\% of its initial mass in Fig.~\ref{fig:illustrative_example}) where isotropic re-emission causes greater contraction of the Roche lobe than would occur in the conservative case for the same amount of mass lost by the donor. In binaries with extended pre-instability evolution (DDI binaries), this prolonged contraction can make non-conservative mass transfer more susceptible to dynamical instability and common-envelope evolution, even though it appears less disruptive early on.

This is particularly relevant for donors with radiative envelopes, for example in the Hertzsprung Gap (HG), for which a mass-transfer instability is preceded by an extended phase of mass loss. During this phase, the orbit and the mass ratio may evolve significantly \citep[e.g.][]{Temmink2025a}, which in turn may alter the Roche lobe's response to mass transfer. In this work, we focus on HG stars, but we expect our results to apply more broadly to all radiative-envelope donors. At solar metallicity, this corresponds roughly to donors with $M_{\rm d}\gtrsim 1.25\,M_\odot$ that overflow their Roche lobe before they reach the base of the giant branch \citep[e.g.][]{Hurley2000}.

\section{Evaluating the accuracy of instantaneous stability assumptions}
\label{sec:param_space}

To assess how a $\beta$-independent $q_{\rm crit}$ and the $\zeta$-method may misrepresent the stability of mass transfer for HG donors, we proceed as follows. We use the MESA code \citep[version \texttt{r12115};][]{Paxton2011, Paxton2013, Paxton2015, Paxton2018, Paxton2019} to re-simulate several systems from \citet{Temmink2025a}. The donors are non-rotating, have $Z=0.02$, the accretor is treated as a point mas, and we impose isotropic re-emission with constant efficiency $\beta \in [0, 0.5, 1]$. The systems span $M_{\rm d,i}=2.5$--$8\,M_\odot$ and donor radii at Roche-lobe overflow $R_{\rm d,RLOF}\simeq 4.6$--$79.1\,R_\odot$ (from the least to the most evolved HG donors in the sample of \citealt{Temmink2025a}), corresponding to $P_{\rm i}\sim 2$--$70$~d for a representative mass ratio $q= 0.25$. We selected these four cases to bracket both donor mass and evolutionary stage at RLOF. Other than the efficiency of the mass transfer, our numerical setup is identical to \citet{Temmink2025a} \footnote{MESA inlists and starting models are available at \href{https://doi.org/10.5281/zenodo.7937496}{doi:10.5281/zenodo.7937496}. To reproduce the accretion efficiencies $\beta$ used here under isotropic re-emission, adjust the \texttt{\&binary\_controls} parameter \texttt{mass\_transfer\_beta} to $1-\beta$: in MESA this denotes the \emph{ejected} fraction from the vicinity of the accretor.}. Appendix~\ref{sec:example_MESA_models} shows the time-dependent MESA evolution for a representative system with the same initial mass ratio as in Fig.~\ref{fig:illustrative_example}. 

We identify unstable mass transfer using the quasi-adiabatic criterion of \citet{Temmink2023a}, which captures the transition to quasi-adiabatic donor evolution preceding runaway mass transfer. For each initial donor mass and radius at RLOF, we use bisection to estimate the value of $q_{\rm crit}$, ensuring that the uncertainty on the critical mass ratio does not exceed 0.025. In the following, we refer to the critical mass ratios obtained via this method as $q_{\rm crit, calc}$. To compare this to what an instantaneous $\zeta$ assumption would predict, we do the following: we assume $\zeta_{\rm d}$ is given by the value of $\zeta_{\rm L}$ at the borderline between stable and unstable mass transfer for our conservative models ($\beta = 1$). Using this value for $\zeta_{\rm d}$, we use Eq.~\ref{eq:zeta_lobe} to determine the critical mass ratio, hereafter $q_{\rm crit, \zeta}$, for a given value for $\beta$ by solving $\zeta_{\rm L}(q,\beta)=\zeta_{\rm d}$ and assuming that the reaction of the star to the loss of mass remains the same (i.e. we assume this $\zeta_{d}$ is independent of $\beta$).

\begin{figure}[tb!]
\centering
\includegraphics[width=0.95\hsize]{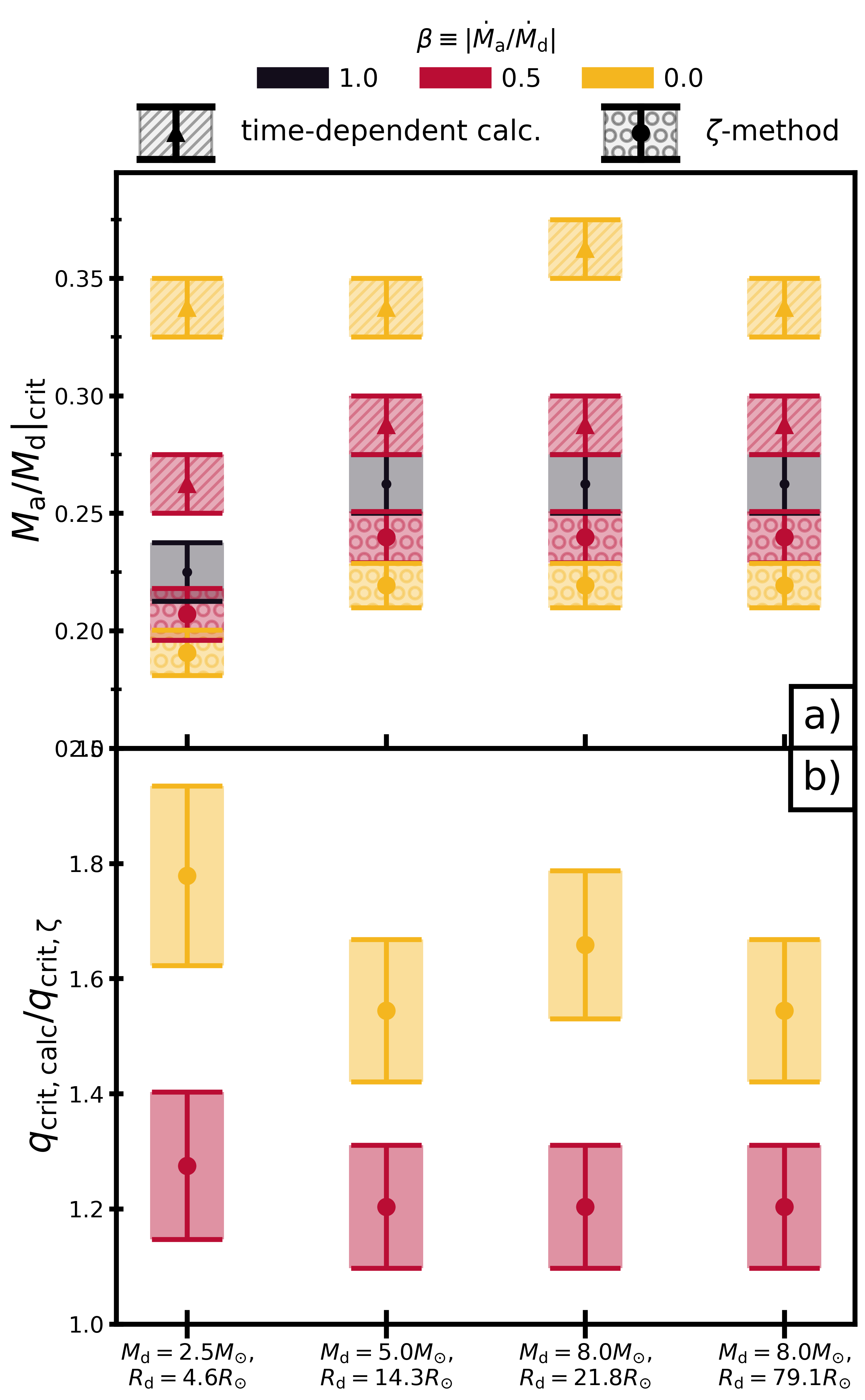}
\caption{Comparison of values for the critical mass ratio for stable mass transfer. The four systems correspond to initial periods of approximately \(P_{\rm i}\simeq 1.8, 7.1, 10.5,\) and \(72.6\) days (for a representative \(q= 0.25\)). Panel a shows values for $q_{\rm crit}$ obtained using the $\zeta$-method ($q_{\rm crit, \zeta}$; sphere hashing) and using MESA calculations ($q_{\rm crit,calc}$; diagonal hashing). For \(\beta=1\) the two methods coincide by construction, since we calibrate \(\zeta_{\rm d}\) from the conservative borderline and then solve \(\zeta_{\rm L}(q,\beta)=\zeta_{\rm d}\) to obtain \(q_{\rm crit,\zeta}\). Panel b shows the ratio of these critical mass ratios \( q_{\rm crit,calc} / q_{\rm crit,\zeta} \). In both panels, the shaded widths are chosen for readability; the vertical extent indicates the uncertainty owing to the finite bisection resolution. The colours correspond to different mass-transfer efficiencies, as shown in the legend.}
\label{fig:representative_zetas_vs_fullcalc}
\end{figure}

The results of this exercise are shown in Fig.~\ref{fig:representative_zetas_vs_fullcalc}a, which shows values for $q_{\rm crit}$ obtained through both methods. The height of the bars represents the uncertainty ranges on the values due to the finite number of bisection steps (see above). Values for $q_{\rm crit}$ obtained using the $\zeta$-method are always more extreme for smaller $\beta$ (i.e. less efficient mass transfer), as expected from the smaller instantaneous $\zeta_{\rm L}$ at smaller $\beta$ (Fig.~\ref{fig:illustrative_example}c; \citealt{Soberman1997}). However, our time-dependent MESA calculations show the opposite trend: $q_{\rm crit,calc}$ increases (i.e. becomes less extreme) as $\beta$ decreases. This is because lower $\beta$ values accelerate the onset of the instability by enhancing orbital contraction early in the mass-transfer phase and hastening the rise of the mass-transfer rate. A similar sign of the $\beta$-dependence of the critical mass ratio was recently reported in an adiabatic stability analysis by \citet{Ge2024}. A stability prescription based on a fixed $q_{\rm crit}$ likewise fails to capture this behaviour, since it cannot account for the decrease in mass-transfer stability with decreasing $\beta$ found in the full calculations. The difference between the prescription-based estimates for $q_{\rm crit}$ and the ones calculated with MESA depends strongly and non-linearly on $\beta$.

Figure~\ref{fig:representative_zetas_vs_fullcalc}b illustrates the ratio $q_{\rm crit,calc}/q_{\rm crit, \zeta}$, quantifying the discrepancy between the full MESA calculations and the $\zeta$-method. For the calibration we have adopted, the $\zeta$-method underestimates the critical mass ratio found from the time-dependent calculations across all cases, and this effect becomes more pronounced as mass transfer becomes less efficient. For $\beta = 0.5$, the full calculations yield critical mass ratios that are 10-40\% higher, and for $\beta = 0$, the discrepancy increases to 40-90\%.

\section{Implications and conclusions}
\label{sec:implications_and_conclusions}

Rapid binary population synthesis codes commonly assess the stability of mass transfer using simplified prescriptions that broadly fall into two categories (as described in Section~\ref{sec:AM_and_DDI}). The instantaneous \(\zeta\)-method attempts to improve on adopting fixed critical mass ratios, as it accounts for changes in the response of the Roche lobe due to the mass-transfer efficiency \(\beta\). As shown in this work, however, as \(\beta\) changes the \(\zeta\)-method predicts changes in mass-transfer stability which are opposite to the behaviour found in time-dependent binary evolution calculations for systems subject to the DDI undergoing isotropic re-emission.  For such systems, time-dependent calculations result in less extreme critical mass ratios for lower \(\beta\) values, while the onset of the instability is accelerated owing to enhanced orbital contraction during the early phases of mass transfer.

As a result, rapid binary population synthesis codes using the \(\zeta\)-method may predict stable mass transfer where a DDI would actually occur, with discrepancies in the critical mass ratio reaching nearly a factor of two for fully non-conservative transfer.  Our findings demonstrate that mass-transfer stability cannot always be reliably assessed through instantaneous methods alone. Instead, accurate modelling requires a time-dependent treatment that accounts for the cumulative effects of mass loss and structural adjustment throughout the evolution of the binary.

\begin{acknowledgements}
We thank the anonymous referee for their constructive remarks and helpful suggestions. KDT acknowledges support from NOVA.
\end{acknowledgements}

\bibliographystyle{aa}
\bibliography{isotropic}

\begin{appendix} 

\section{Orbital evolution under isotropic re-emission}
\label{sec:turning_point}

The orbital angular momentum of a circular binary is given by
\begin{equation}
J_{\rm orb}
=
M_{\rm d}M_{\rm a}
\left(\frac{Ga}{M_{\rm d}+M_{\rm a}}\right)^{1/2}.
\end{equation}
Taking a logarithmic time derivative gives
\begin{equation}
\label{eq:adot_general}
\frac{\dot a}{a}
=
2\frac{\dot J_{\rm orb}}{J_{\rm orb}}
-2\frac{\dot M_{\rm d}}{M_{\rm d}}
-2\frac{\dot M_{\rm a}}{M_{\rm a}}
+\frac{\dot M_{\rm d}+\dot M_{\rm a}}{M_{\rm d}+M_{\rm a}}.
\end{equation}

Under the assumption of isotropic re-emission, a fraction \(\beta\) of the transferred mass is accreted by the companion, so that
\(\dot M_{\rm a}=-\beta\dot M_{\rm d}\), while the remaining fraction \(1-\beta\) is expelled from the vicinity of the accretor carrying the accretor's specific orbital angular momentum. Writing the mass ratio as \(q\equiv M_{\rm a}/M_{\rm d}\), this implies
\begin{equation}
\frac{\dot J_{\rm orb}}{J_{\rm orb}}
=
(1-\beta)\frac{\dot M_{\rm d}}{M_{\rm d}}\frac{1}{q(1+q)}.
\end{equation}
Substituting these relations into Eq.~\eqref{eq:adot_general} gives
\begin{equation}
\label{eq:dadot_q_simplified}
\frac{\dot a}{a}
=
-2\frac{\dot M_{\rm d}}{M_{\rm d}}
\left[
1-\frac{\beta}{q}
-\frac{(1-\beta)(2+q)}{2q(1+q)}
\right].
\end{equation}

During Roche-lobe overflow the donor loses mass, so \(\dot M_{\rm d}<0\), and the prefactor
\(-2\,\dot M_{\rm d}/M_{\rm d}\) is strictly positive. It follows that the sign of \(\dot a/a\) is set by the terms in the bracket of Eq.~\eqref{eq:dadot_q_simplified}. Setting this term equal to zero leads to a quadratic equation in \(q\),
\begin{equation}
2q^{2}+(1-\beta)q-2=0,
\end{equation}
of which the only physical (non-negative) solution is 
\begin{equation}
\label{eq:qturn}
q_{\rm turn}
=
\frac{\beta-1+\sqrt{(1-\beta)^{2}+16}}{4}.
\end{equation}

We note that the turning-point mass ratio \(q_{\rm turn}\) is $\leq 1$ and decreases with decreasing \(\beta\). Differentiating Eq.~\eqref{eq:qturn} gives
\begin{equation}
\frac{{\rm d}q_{\rm turn}}{{\rm d}\beta}
=\frac{1}{4}\left(1+\frac{\beta-1}{\sqrt{(\beta-1)^2+16}}\right) > 0
\qquad \forall \; \beta \in [0,1].
\end{equation}
Hence \(q_{\rm turn}\) monotonically increases with \(\beta\), and is thus lower for less efficient (lower \(\beta\)) mass transfer. This means that when \(\beta < 1\), the orbit starts expanding when the donor is still more massive than the accretor.

\section{Example MESA models}
\label{sec:example_MESA_models}

\begin{figure}[]
\centering
\includegraphics[width=\hsize]{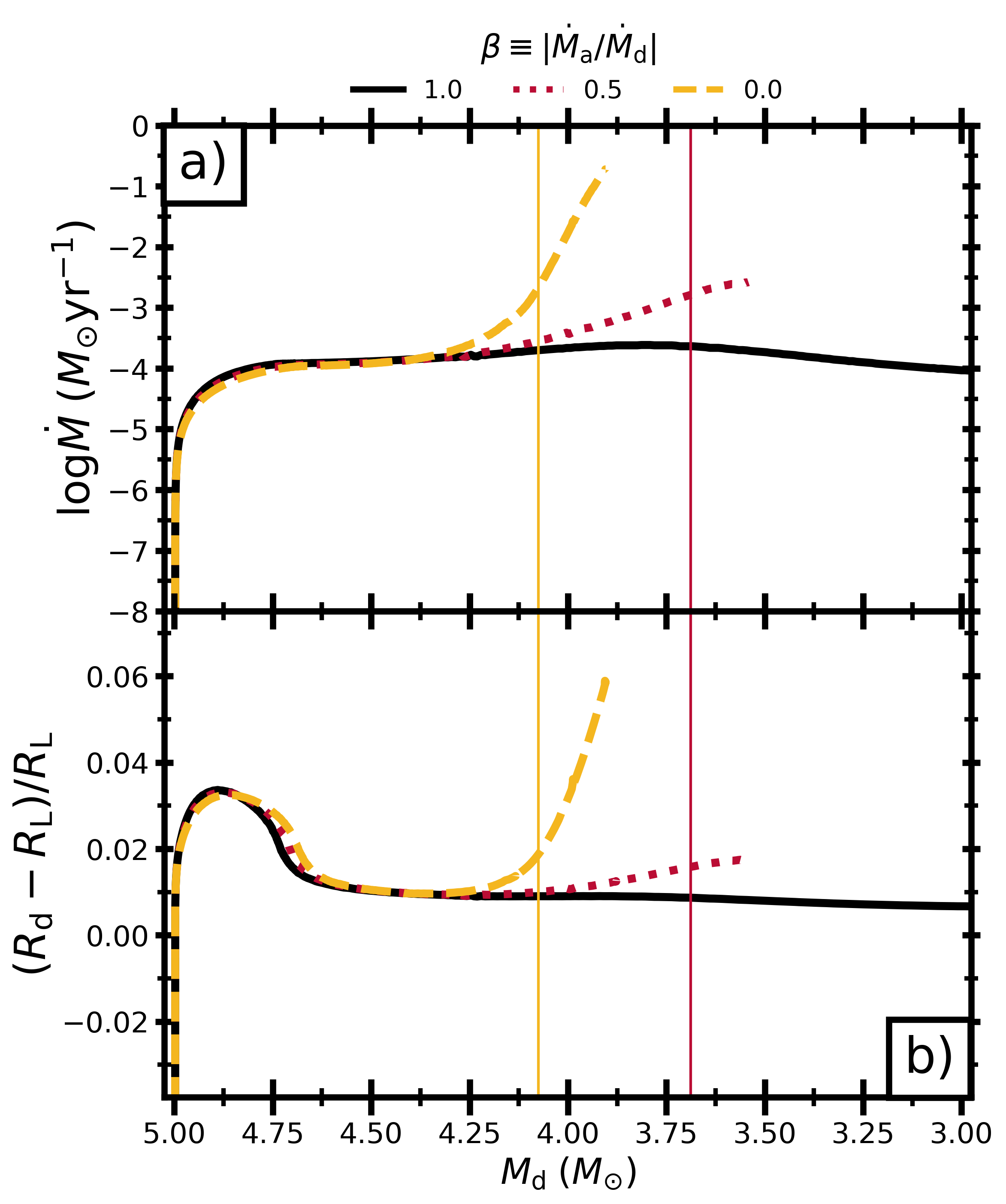}
\caption{Evolution of the mass transfer rate (panel a) and relative Roche lobe overflow (panel b) for a representative \(5\,M_{\odot}\) donor star losing mass to an initially \(1.375\,M_{\odot}\) companion with an initial period of about 7 days. The colours and styles of the curves correspond to different mass-transfer efficiencies, as indicated in the legend. Vertical lines indicate the donor-mass at which the quasi-adiabatic instability criterion \citep[see][]{Temmink2023a} is reached in our MESA simulations (with numerical setup as in Section~\ref{sec:param_space}). Mass transfer is stable in the fully conservative case.}
\label{fig:illustrative_example_MESA}
\end{figure}

To illustrate the time-dependent evolution that underlies the stability boundaries discussed in Section~\ref{sec:param_space}, Figure~\ref{fig:illustrative_example_MESA} shows a representative \texttt{MESA} model with the same initial mass ratio as in Figure~\ref{fig:illustrative_example} ($M_{\rm d,i}=5\,M_\odot$, $M_{\rm a,i}=1.375\,M_\odot$; $P_{\rm i}\simeq 7$~d). 

The vertical lines (which are also shown in Figure~\ref{fig:illustrative_example}) mark the donor-mass coordinate at which the quasi-adiabatic instability criterion of \citet{Temmink2023a} is met. For this system, reaching the criterion coincides with the mass-transfer rate rising to $\dot{M}\sim10^{-3}\,M_\odot\,{\rm yr^{-1}}$ and with rapidly increasing overfill of the Roche lobe. While the fully conservative sequence remains stable, decreasing the mass-transfer efficiency leads to a more rapid approach to the instability, both in terms of time as well as the amount of donor mass lost.

\end{appendix}

\end{document}